\documentclass[twocolumn]{aastex6}
\usepackage{multirow}
\usepackage{units}

\def\be{\begin{eqnarray}}
\def\ee{\end{eqnarray}}

\begin{document}

\title{Is GRB 110715A the progenitor of FRB 171209?}
\author{Xiang-Gao Wang\altaffilmark{1,2}, Long Li\altaffilmark{1,2,3}, Yuan-Pei Yang\altaffilmark{4}, Jia-Wei Luo\altaffilmark{5}, Bing Zhang\altaffilmark{5}, Da-Bin Lin\altaffilmark{1,2}, En-Wei Liang\altaffilmark{1,2}, Song-Mei Qin\altaffilmark{6}}
\altaffiltext{1}{Guangxi Key Laboratory for Relativistic Astrophysics, School of Physical Science and Technology, Guangxi University, Nanning 530004, China; wangxg@gxu.edu.cn}
\altaffiltext{2}{GXU-NAOC Center for Astrophysics and Space Sciences, Nanning 530004, China}
\altaffiltext{3}{School of Astronomy and Space Science, Nanjing University, Nanjing 210093, China}
\altaffiltext{4}{South-Western Institute for Astronomy Research, Yunnan University, Kunming 650500, China; ypyang@ynu.edu.cn}
\altaffiltext{5}{Department of Physics and Astronomy, University of Nevada Las Vegas, NV 89154, USA; zhang@physics.unlv.edu}
\altaffiltext{6}{Mathematics and Physics Section, Guangxi University of Chinese Medicine, Nanning 53001,China}

\begin{abstract}
The physical origin of fast radio bursts (FRBs) is unknown.
Young magnetars born from gamma-ray bursts (GRBs) have been suggested to be a possible central engine of FRBs.
We test such a hypothesis by systematically searching for GRB-FRB spatial associations from 110 FRBs and 1440 GRBs.
We find that one FRB event, FRB 171209, discovered by the Parkes telescope is spatially coincident with a historical long-duration GRB 110715A at $z=0.82$. The afterglow of GRB 110715A is consistent with being powered by a millisecond magnetar. The extragalactic dispersion measure of FRB 171209 is in excess of that contributed by the intergalactic medium, which can be interpreted as being contributed by a young supernova remnant associated with the GRB. Overall, the significance of the association is $(2.28 - 2.55) \sigma$.  If the association is indeed physical, our result suggests that the magnetars associated with long GRBs can be the progenitors of at least some FRBs.
\end{abstract}

\keywords{star: gamma-ray burst - star: fast radio burst - star: magnetar - radiation mechanisms: non-thermal }

\section{Introduction}
\label{sec:Intr}
Fast radio bursts (FRBs) are mysterious radio transients with millisecond durations, extremely high brightness temperatures and large dispersion measures (DMs) \citep[e.g.,][]{lorimer07,thornton13,petroff19,cordes19}. Their DMs are in excess of the Galactic contribution and the precise localizations of the host galaxies of a few FRBs suggest that they are extragalactic  \citep[e.g.,][]{lorimer07,thornton13,chatterjee17,bannister19,Prochaska19,ravi19,marcote20}.
A persistent radio emission with luminosity of $L\sim10^{39}~{\rm erg~s^{-1}}$ at a few GHz was discovered to be spatially coincident with FRB 121102, which showed a non-thermal spectrum that deviates from a single power-law spectrum from 1 GHz to 26 GHz \citep{chatterjee17}. One possibility is that such a persistent radio emission source  originates from a shocked nebula associated with a young magnetar born in a supernova (SN) or a gamma-ray burst (GRB)  \citep{murase16,metzger17}. On the other hand, there is no confirmed multi-wavelength transient being associated with any FRB \citep[e.g.,][]{petroff15,cal16,zhangbbfrb17,MAGIC18,tin19}. There might be three main reasons: 1. The fluxes of the multi-wavelength counterparts of FRBs are low, e.g., for typical parameters, the FRB afterglows are very faint \citep{yi14}; 2. The duration of the multi-wavelength transient may be shorter than the time resolution of a detector, e.g., the prompt high-energy emission associated with the FRB itself \citep{yang2019}; 3. the time delay between the multi-wavelength transient and the FRB is longer than the observation time, e.g., FRBs emitted from a young magnetar born from a catastrophic event (such as a GRB or a SN) may have a long delay with respect to the event itself \citep{murase16,metzger17}.

Lacking confirmed multi-wavelength transients associated with FRBs,
the physical origin of FRBs is still unknown. The current FRB models can be divided into two categories\footnote{A complete list of FRB progenitor models can be seen in \citet{platts19}.}: catastrophic models \citep[e.g.,][]{kashiyama13,totani13,falcke14,zhang14,zhang16a,wangjs16} and non-catastrophic models \citep[e.g.,][]{murase16,metzger17,zhang17,zhang20,dai16,margalitmetzger18,wangfy20,iok20}. The former suggested that an FRB is directly associated with a catastrophic event, and the time delay between the FRB and the catastrophic event is short. The latter usually involved a compact star, e.g., a neutron star or a black hole, that was born in a catastrophic event as the progenitor of an FRB. Since the compact star can exist for a much longer time, the time delay between the FRB and the catastrophic event is allowed to be relatively long.

GRBs are the most luminous catastrophic events, and are produced by core collapse of massive stars or binary compact star mergers \citep{mes06,zhang18book}. Although GRBs are much rarer than FRBs, the following reasons have been raised to suggest that a fraction of FRBs could be associated with GRBs: 1. an FRB might occur when a supermassive magnetar born in a GRB collapses to a black hole, so called ``blitzar'' scenario \citep{falcke14,zhang14}; 2. an FRB might be related to the merger of a binary neutron star which produces a short GRB \citep{totani13,wangjs16}; 3. a GRB as the source of astrophysical stream could interact with the magnetosphere of a neutron star to produce an FRB \citep{zhang17}; 4. a GRB could produce a young magnetar that emits FRBs at a much later epoch \citep[e.g.,][]{metzger17,margalit19,wangfy20}. In the first three scenarios, an FRB could occur from milliseconds before to a few thousand seconds after the GRB. The fourth scenario allows a much longer delay of FRBs with respect to the GRB. Related to this, recently \citet{eft19} discovered a radio source coincident with the SLSN PTF10hgi, similar to the persistent radio emission of FRB 121102 \citep{chatterjee17}, about 7.5 years post-explosion, which might be emitted by the magnetar born in the SLSN. However, no FRB was detected from the source yet.

In general, the searches for GRB-FRB associations have so far given no confirmed results \citep[e.g.,][]{bannister12,palaniswamy14,delaunay16,scholz16,yamasaki16,zhangbbfrb17,xi17,guidorzi19,yangyh19,cunningham19,Tavani2020ATel}. In particular, \citet{menyp19} recently performed dedicated observations of the remnants of six GRBs with evidence of having a magnetar central engine, but these observations did not lead to detection of any FRB from these remnants during a total of $\sim20$h of observations.

In this work, we adopt a different approach to test the hypothesis that GRBs can be the progenitor of FRBs. We systematically search for GRB-FRB association events based on the precise localization of GRB afterglows, allowing a few years of time delay between a GRB and an FRB.
Observationally, GRBs are typically localized  by the \emph{Neil Gehrels Swift Observtory}, i.e. the burst is detected by \emph{Swift}/BAT and quickly localized by \emph{Swift}/XRT with a several-arcsecond error bar, and later further localized by \emph{Swift}/UVOT or groundbased telescopes to sub-arcsecond precision.
Based on the archival \emph{Swift}/XRT and optical observational data, we search for possible GRB-FRB spatial association candidates. We detect one possible association between FRB 171209 \citep{oslowski19} and GRB 110715A at $z=0.82$ \citep{sanchez-ramirez17}. This paper is organized as follows. In Section 2, we present the search method and result. The GRB 110715A - FRB 171209 association is discussed in Section 3 in detail. The results are summarized with discussion in Section 4.
Throughout the paper, we adopted a concordance cosmology with parameters $H_{0}= 71$ ${\rm km s^{-1}}$ ${\rm Mpc^{-1}}$, $\Omega_{M}=0.30$, $\Omega_{\Lambda}=0.70$, and temporal and spectral slopes of GRB afterglow emission are defined as $F\propto t^{-\alpha} \nu^{-\beta}$. Moreover the convention $Q = 10^nQ_n$ is adopted in cgs units.

\section{Search for Gamma-Ray Bursts Associated with Fast Radio Bursts in Archival \emph{Swift}/XRT and optical observational data}
\label{sec:fit}
\begin{figure}[t]
\center
\begin{tabular}{c}
\includegraphics[angle=0,width=0.4\textwidth]{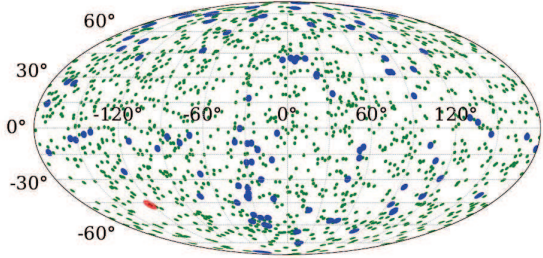}
\end{tabular}
\caption{Sky celestial coordinate distributions of 110 FRBs and 1440 GRBs in our sample. The FRBs and GRBs are marked with blue and green circles, respectively. The positions of GRB 110715A and FRB 171209 are highlighted with a red circle.}
\label{fig:location}
\end{figure}

\begin{figure}[t]
\center
\begin{tabular}{c}
\includegraphics[angle=0,width=0.4\textwidth]{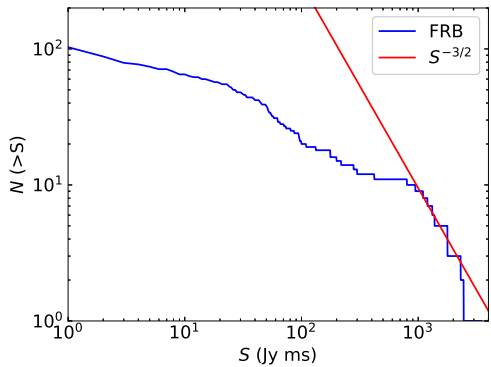}
\end{tabular}
\caption{The $\log N - \log S$
distribution of the FRBs in our sample.}
\label{fig:logSN}
\end{figure}

Since the discovery of the first FRB \citep[][]{lorimer07}, 110 FRBs have been reported in the literature as of February 2020\footnote{http://www.frbcat.org\citep{petroff16}.} \citep[e.g.,][]{petroff15,chime19,casentini19}. Meanwhile, up to now there are 1440 GRBs that have afterglow detections, including 845 GRBs with optical detections and 595 with \emph{Swift}/XRT\footnote{https://swift.gsfc.nas.gov/archive/grb\_table/} detections only. Among them, the numbers of long and short duration GRBs are 1320 and 120, respectively.  Figure \ref{fig:location} shows the sky distribution of our samples (110 FRBs and 1440 GRBs) in celestial coordinates. The GRBs in our sample show a large-scale isotropic distribution, which is well known from the BATSE observations \citep[][]{briggs96}. Although the sample size of FRBs is smaller than that of GRBs, the FRBs in our sample also show an isotropic distribution, consistent with their cosmological origin. As shown in Figure \ref{fig:logSN}, the distributions of the FRB fluence ($\log N - \log S$)\footnote{The fluence ($S$) values are obtained from http://www.frbcat.org.} show a tendency with $N(>S)\propto S^{\rm -3/2}$ at high $S$ values. The deviation from the 3/2 power law is evident at low $S$ values, which may be related to the spatial inhomogeneity effect and likely also observational biases and instrumental effects.

We perform a systematic search for GRBs that satisfy the following three criteria: (1) the GRB position is consistent with that of an FRB; (2) the GRB occurred earlier than the FRB if a position coincidence is discovered; (3) the redshift of the GRB is lower than the maximum FRB redshift derived from its DM.

We found only one GRB that is located at the position of an FRB, i.e. the GRB 110715A - FRB 171209 spatial association. GRB 110715A was triggered by the \emph{Swift}/BAT on 2011 July 15 (UT dates are adopted) at $13:13:50$ ($T_0$), with $T_{90} = 13$~s \citep{sonbas110715AGCN,ukwatta110715AGCN}. It was also detected by the {\it Konus-wind} \citep{golenetskii110715AGCN}. The \emph{Swift}/XRT and \emph{Swift}/UVOT began observing its X-ray and optical afterglows at 90.9~s and 99~s after the BAT trigger, respectively \citep{kuin110715AGCN,evans110715AGCN}. It was also followed up by ground-based optical telescopes GROND \citep{updike110715AGCN} and AAVSO \citep{nelson110715AGCN}; submm telescopes APEX \citep{postigo110715AGCN} and ALMA \citep{sanchez-ramirez17}; and the radio telescope ATCA \citep{hancock110715AGCN}. The position of GRB 110715A, defined by its optical afterglow, is (RA$_{\rm J2000.0}$, DEC$_{\rm J2000.0}$) = (15$^{\rm h}$50$^{\rm m}$44$^{\rm s}$.09, $-46^\circ$14$^\prime$06$^{\prime\prime}$.53), with an estimated uncertainty of 0.56 arc sec (radius, 90\% confidence) \citep{kuin110715AGCN}. The redshift of GRB 110715A was measured to be $z = 0.82$ \citep{piranomonte110715AGCN}.

On the other hand, FRB 171209 \citep{oslowski19} was detected 2338 days ($\sim$6.4 yr) after the GRB 110715A trigger. It was the first FRB detected as part of the commensal search during PPTA observations, with a position (RA$_{\rm J2000.0}$, DEC$_{\rm J2000.0}$) = (15$^{\rm h}$50$^{\rm m}$25$^{\rm s}$, $-46^\circ$10$^\prime$20$^{\prime\prime})$, with an uncertainty of 7.5 arcmin (radius, 2.355$\sigma$ confidence). The DM value is $1457.4 \pm 0.03$ cm$^{-3}$ pc and the DM value contribution from the Milky Way is $\rm DM_{gal}=235$ cm$^{-3}$ pc \citep{oslowski19}. Using a simple DM-$z$ relation ${\rm DM_{IGM}} \sim 855{\rm pc~cm^{-3}} z$ \citep{zhang18}, one can estimate
the maximum redshift of FRB 171209, which is $\sim ({\rm DM-DM_{gal}})/855{\rm pc~cm^{-3}} = 1.43$. Due to the existence of large-scale structures, the uncertainty of the DM contributed by the IGM is about $\sigma_{\rm IGM}\sim300~{\rm pc~cm^{-3}}$.
Thus the maximum redshift is constrained in the range of $z<(1.08-1.78)$. This is larger than the redshift of GRB 110715A.

To calculate of chance possibility for the putative GRB 110715A - FRB 171209 association, we assume that the spatial distribution of GRBs is isotropic and the number of GRBs within a specific sky area and time interval satisfies the Poisson distribution. The chance probability of having at least one GRB in the error circle of one FRB can then be written as
\be
 P_{1} = 1 - \lambda^0\exp(-\lambda)/0! = 1 - \exp(-\lambda),
\ee
where $\lambda = \rho S$ is the expected number of GRBs in the FRB error region $S$. The surface number density of GRBs is $\rho\approx 1440/41252.96\approx0.035/\rm deg^2$. For a circular region with a radius  $\delta R$ (in unit of deg), we can derive its area $S \approx [41252.96(1-\cos\delta R)]/2$.

To estimate the p-value of the chance coincidence, we adopt two approaches. First, for a conservative estimate, we use the uncertainty of 7.5 arcmin defined by the error bar of the FRB position, i.e. 
$\delta R = 0.125^\circ$. We obtain the chance probability of having at least one (out of 1440) GRB whose distance to FRB 171209 is smaller than $0.125^\circ$, which gives $P_{1}\approx0.0017$. The chance probability of having only one such association for all 110 FRBs can be estimated as $P=1-(1-P_1)^{110} \approx 17.1\%$. We verify this simple estimate through Monte Carlo simulations. We randomly generate 1440 GRBs and 110 FRBs in the sky. Based on $10^5$ simulations, the chance probability of having a GRB/FRB pair with a separation smaller than $0.125^\circ$ is $17.4\%$, consistent with the analytical estimate.

One also needs to consider two other criteria for an association, i.e. the timing criterion (the GRB needs to occur before the FRB) and the redshift criterion (the maximum redshift derived from the FRB DM is larger than that of the GRB). To do this, we use the observed distributions of the detection time and redshift for both GRBs and FRBs to perform the simulations. Since most GRBs were detected earlier than FRBs (FRBs were discovered much later than GRBs), adding the timing criterion does not reduce the chance probability significantly, i.e. $\sim$ 14.1\%. However, since the average redshift of GRBs is higher than the average maximum redshift of FRBs, adding the redshift criterion reduces the chance probability significantly to $\sim$ 2.3\%, which corresponds to a significance of 2.28$\sigma$.

Second, since GRB 110715A is well located inside the error circle of FRB 171209,one may use the angular distance between the centers of the error boxes of the two events, $0.0836^\circ$, as $\delta R$\footnote{This approach was adopted by the IceCube team to claim a possible association between the neutrino trigger event IceCube-170922A and the blazar TXS 0506+056 \citep{Icecube}.}
One can obtain the chance probability of having at least one (out of 1440) GRB whose distance to FRB 171209 is smaller than $0.0836^\circ$, i.e. $P_{1}\approx0.0007$. We also randomly generate 1440 GRBs and 110 FRBs in the sky. Based on $10^5$ simulations, the chance probability of having one GRB/FRB pair with a separation smaller than $0.0836^\circ$ is $7.6\%$. Considering the timing criterion, we obtain $P=6.3\%$. When the redshift criterion is also considered, the final chance probability is 1.1\%, which corresponds to a 2.55$\sigma$ confidence level for the GRB 110715A/FRB 171209 association.

\section{Is GRB 110715A associated with FRB 171209?}
\label{sec:dis}

Even though statistically one cannot establish a firm association between GRB 110715A and FRB 171209, it is nonetheless interesting to investigate whether physically such an association makes sense.

\subsection{Magnetar as central engine of GRB 110715A}
\begin{figure}[t]
\center
\begin{tabular}{c}
\includegraphics[angle=0,width=0.4\textwidth]{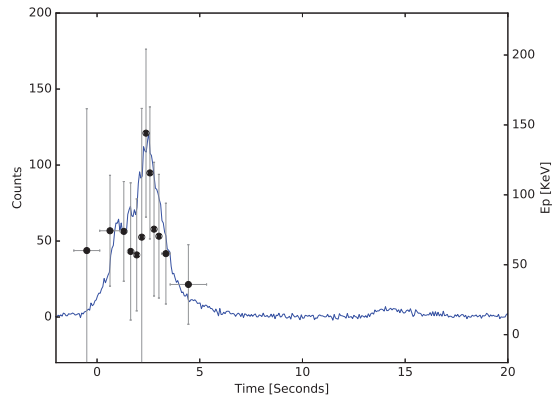}
\end{tabular}
\caption{The BAT lightcurve (blue line) and its $E_p$ (black circle) evolution of GRB 110715A. The isotropic $\gamma$-ray energy is $E_{\rm \gamma, iso}=1.06 \pm 0.10 \times 10^{53}$ erg in the 1-10$^4$ keV band.}
\label{fig:BAT}
\end{figure}

\begin{figure}[t]
\center
\begin{tabular}{c}
\includegraphics[angle=0,width=0.4\textwidth]{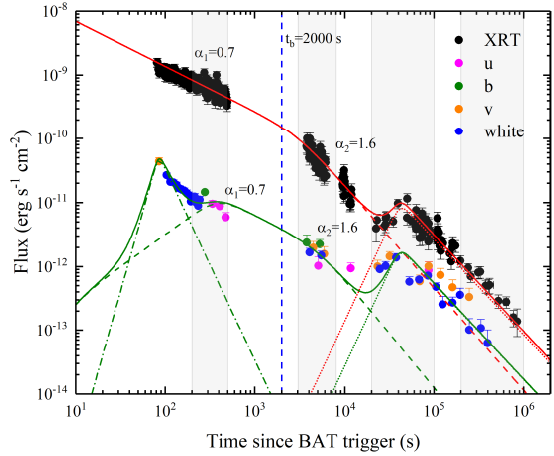}
\end{tabular}
\caption{Lightcurves of X-ray and optical afterglows of GRB 110715A. The light curves are decomposed into multiple components (dashed or dash-dotted lines). The solid lines represent the best fit to the data. The vertical blue dashed lines mark the break time between the shallow decay phase to the normal decay phase. The grey zones represent the time slices for the afterglow SED analysis.  }
\label{fig:lc}
\end{figure}

The \emph{Swift}/BAT time-integrated spectrum of GRB~110715A can be well fitted with a Band function \citep{band93}, with $E_{\rm peak} = 92.8 \pm 18.1$ keV, $\alpha = -1.23 \pm 0.12$, and $\beta = -2.05 \pm 0.19$ and $\chi^2$=0.98 (as shown in Figure \ref{fig:BAT}). We obtained the isotropic $\gamma$-ray energy $E_{\rm \gamma, iso}=1.06 \pm 0.10 \times 10^{53}$ erg in the 1-10$^4$ keV band. The results from the time-resolved spectral analysis show the ``flux-tracking'' pattern for  $E_p$.
To fit the GRB 110715A afterglow lightcurves, we employed
a broken power-law function
\begin{equation} F=F_1\left [
\left (   \frac{t}{t_b}\right)^{\omega\alpha_1}+\left (
\frac{t}{t_b}\right)^{\omega\alpha_2}\right]^{1/\omega},
\end{equation}
\noindent
where $F_1$ is the flux normalization, $\alpha_{1}$ and $\alpha_{2}$ are the afterglow flux decay indices before and after the break time ($t_b$), respectively, and $\omega$ is a smoothness parameter which represents the sharpness of the break. Figure \ref{fig:lc} shows the X-ray and optical light curves of GRB110715A. The X-ray light curve can be well fitted by a broken power-law function, with the best-fit power-law slope $\alpha_{\rm X,1}=0.70^{+0.04}_{-0.05}$ (shallow decay) before the break ($t_{\rm b}=T_0+2.0^{+0.4}_{-0.3}$~ks) and $\alpha_{\rm X,2}=1.60^{+0.11}_{-0.09}$ (normal decay) after the break, respectively. There is a re-brightening component appearing at $\sim T_0+50^{+0.4}_{-0.3}$~ks. For the optical light curve, there is an early steep decay phase, which may be interpreted as the reverse shock emission as the ejecta is decelerated. This is followed by a shallow decay phase (with $\alpha_{\rm O,1}=0.70^{+0.13}_{-0.12}$) breaking at $t_{\rm b}$ and further decays with $\alpha_{\rm O,2}=1.60^{+0.15}_{-0.11}$.
The re-brightening component also appeared in the optical afterglow. The result of the temporal analysis suggests that the X-ray and optical afterglow show an achromatic behavior \citep{Wang15}.

We also analyze the spectral energy distributions (SEDs) of GRB 110715A afterglow, by jointly fitting the optical and XRT data with the Xspec package \citep{arnaud96} and the optical data that are corrected for Galactic extinction based on the burst direction, with $A_{\rm V} = 0.030, A_{\rm R} = 0.119$ and $A_{\rm I} = 0.016$ \citep{schlafly11}. The extinction 
in the host galaxy is also taken into account
assuming an extinction curve similar to that of Small Magellanic Cloud (SMC) with its Standard value of the ratio of total to selective extinction $R_{\rm v,SMC} = 2.93$ \citep{pei92}. The equivalent hydrogen column density of our Galaxy is $N_{\rm H} = 4.33 \times 10^{21}$ $\rm cm^{-2}$. The equivalent hydrogen column density of the host galaxy $N_{\rm H}^{\rm host} = (4.22 \pm 2.95) \times 10^{21}$ $\rm cm^{-2}$ is derived from the time integrated XRT spectrum. We fix these values in our time-resolved spectral fits. We subdivided the broadband data into four temporal ranges (as
marked in Figure \ref{fig:lc}). The SEDs of the joint optical and X-ray spectra can be well fitted with a single absorbed power-law function. The photon indices $\Gamma$ (the spectral index $\beta=\Gamma-1$) are 1.69, 1.70, 1.87 and 1.89 for the Slice 1 ($T_0$ + [200,~500]~s), Slice 2 ($T_0$ + [$3\times10^3$,$8\times10^3$]~s), Slice 3 ($T_0$ + [$2\times10^4$,$1\times10^5$]~s), and Slice 4 ($T_0$ + [$2\times10^5$,$1\times10^6$]~s), respectively. There is no obvious spectral evolution observed in the afterglow phase. The temporal slopes of the normal decay phase ($\alpha_{\rm X,II}$ and $\alpha_{\rm O,II}$) are well consistent with the closure relations ($\alpha-\beta$) of the fireball external shock model $\alpha= 3\beta /2+0.5=1.54\pm0.08$, which are located in spectral regime ($\nu_{\rm m}<\nu<\nu_{\rm c}$) in the wind stellar medium \citep[e.g.][]{gao13}.
For the shallow decay phase closure relation $\alpha=q/2+(2+q)\beta/2$ \citep{zhang06}, we obtained the energy injection parameter $q=0$ for $\alpha_{\rm X,I}=0.70^{+0.04}_{-0.05}$ and $\alpha_{\rm O,I}=0.70^{+0.13}_{-0.12}$, which is consistent with the energy injection from the spin-down of a millisecond magnetar \citep{dai98,zha01}. \cite{cikintoglu19} also argued that the millisecond magnetar could be the central engine of GRB 110715A.

\begin{figure}
\center
\begin{tabular}{c}
\includegraphics[angle=0,width=0.4\textwidth]{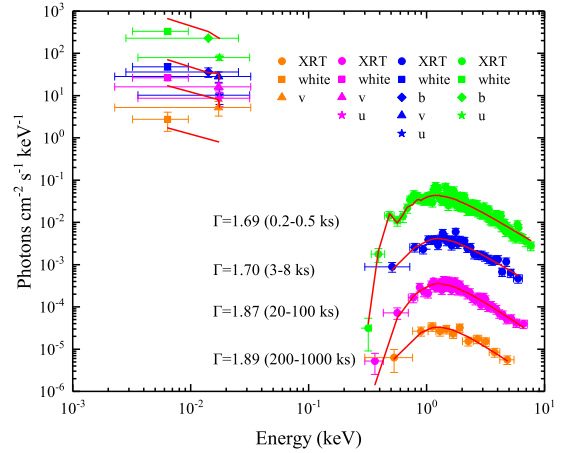}
\end{tabular}
\caption{The SED analysis of GRB 110715A. Joint spectral fits of the X-ray and optical afterglows in four selected time intervals. The solid lines show the intrinsic power-law spectra derived from the joint fits. Different spectral bands are denoted in different symbols: XRT data (circle), white band (square), b-band (prismatic), v-band (triangle), and u-band (star). The photon indices $\Gamma$ in different time intervals  are also marked in different colors.}
\label{fig:SED}
\end{figure}

We further investigate the afterglow data with the standard forward shock model with energy injection ($q=0$).
A Markov Chain Monte Carlo (MCMC) method is adopted to search for the best fitting parameters. The results are shown in Figure 4. One can see that the model can well reproduce the data. The best fitting parameters are: the isotropic kinetic energy $E_{\rm K,iso}=2\times 10^{53}$ erg, the initial Lorentz factor $\Gamma_{\rm 0}=45$, the fraction of shock energy to electrons  $\epsilon_{\rm e}=0.268$, the fraction of shock energy to magnetic fields $\epsilon_{\rm B}=1.1\times 10^{-6}$, wind density parameter $A_\ast=0.25$, the energy injection luminosity $L_{\rm 0}=1\times 10^{50} $ erg s$^{-1}$, and the duration of energy injection $t_b = 2000$ s. The fitting parameters are consistent with the statistical properties of a large sample of GRBs \citep[e.g.,][]{Wang15}.

Since the energy injection $q=0$ is well consistent with the magnetar spin-down model, we can  derive the magnetar parameters of GRB 110715A based on the data. The maximum energy is the total rotational energy of a millisecond magnetar and is defined as
\begin{equation}
E_{\rm rot} = \frac{1}{2} I \Omega_{0}^{2}
\simeq 2 \times 10^{52}~{\rm erg}~
M_{1.4} R_6^2 P_{0,-3}^{-2},
\label{Erot}
\end{equation}
where $I$ is the moment of inertia, $P_0$ is the initial spin period, $\Omega_0 = 2\pi/P_0$ is the initial angular frequency of the neutron star, $M_{1.4} = M/1.4M_\odot$, and $R$ is the radius of the magnetar. The isotropic $\gamma$-ray and kinetic energies are larger than this value, suggesting that the outflow is beamed, with a beaming factor $f_b = 1-\cos\theta_j < 0.1$, where $\theta_j$ is the half opening angle of the jet. Based on the characteristic spin down luminosity $L_0$ and the spindown timescale  $\tau$ of a magnetar as shown in Equation (6) and (8) in \cite{zha01}, one can calculate the surface polar cap magnetic field strength $B_p$ and the initial spin period $P_0$:

\be
B_{\rm p,15} & = & 2.05(I_{45} R_6^{-3} L_{0,49}^{-1/2} \tau_{3}^{-1})~\rm G, \\
P_{0,-3} & = & 1.42(I_{45}^{1/2} L_{0,49}^{-1/2} \tau_{3}^{-1/2})~\rm s.
\ee
Observationally, the spindown luminosity $L_0$ can be generally written as \citep{LuZhang14}
\be
L_0&=&[L_{\rm X,iso}+E_{\rm K,iso}(1+z)/t_b]f_b,
\ee
where $L_{\rm X,iso}$ is the X-ray luminosity due to internal dissipation
of the magnetar wind, which is negligible in our case.

Since no jet break is observed in GRB 110715A, we can use the epoch of the last observational data point to set a lower limit on $\theta_{j}$ \citep{wang18}, i.e.
$\theta_{j} > 6.2^{\rm o}$. Using $E_{\rm K,iso}=2\times 10^{53}$ erg, $L_{\rm X,iso}=3.28\times10^{47}$ erg s$^{-1}$, and $\tau=t_{b}/(1+z)=2000/(1+0.82)=1099$ s, we obtain
$P_0 <  3.59 $ ms and $B_{\rm p} < 4.95 \times 10^{15}$ G, respectively.
These parameters fall into the regime of typical young magnetars for GRB central engines. Such a magnetar is believed to power repeating FRBs when the environment becomes clean \citep{murase16,metzger17,margalitmetzger18}.

\subsection{Is the magnetar the progenitor of FRB 171209?}
\label{sec:FRBmag}

As reported by \citet{oslowski19}, FRB 171209 has a duration of $\Delta t\sim0.138~{\rm ms}$ and a fluence of $f_\nu\gtrsim3.7~{\rm Jy~ms}$ at $\nu\sim1~{\rm GHz}$. If FRB 171209 is indeed associated with GRB 110715A, according to the redshift $z=0.82$ of GRB 110715A, the luminosity distance is $d_{\rm L}\simeq5~{\rm Gpc}$. The isotropic energy of FRB 171209 is about $E_{\rm FRB}\sim4\pi d_{\rm L}^2\nu f_\nu\gtrsim 1.1\times10^{41}~{\rm erg}$. If this energy is provided by the magnetic energy of the underlying magnetar, one may place a most demanding constraint on the strength of the magnetic field of the underlying magnetar assuming isotropic FRB radiation.
The emission radius can be approximately estimated as $r_e\sim c\Delta t\simeq4.1\times10^6~{\rm cm}$. The magnetic field strength at $r_e$ should satisfy
\be
\frac{B^2}{8\pi}\left(\frac{4\pi}{3}r_e^3\right)\gtrsim E_{\rm FRB},
\ee
where $B=B_{\rm p}(r_e/R)^{-3}$. Therefore, the observation of FRB 171209 demands that the surface polar cap magnetic field strength is
\be
B_{\rm p}\gtrsim\left(\frac{6Er_e^3}{R^6}\right)^{1/2}\simeq6.8\times10^{12}~{\rm G},
\ee
which is consistent with the observation constraints derived from  the afterglow emission of GRB 110715A.

According to the redshift $z=0.82$ of GRB 110715A, the DM contribution from the IGM is given by \citep{zhang18},
\be
{\rm DM_{IGM}}\simeq855~{\rm pc~cm^{-3}}z\simeq700~{\rm pc~cm^{-3}},
\ee
and the local DM from the host galaxy is
\be
{\rm DM_{host}}=(1+z)({\rm DM_E}-{\rm DM_{IGM}})\simeq950~{\rm pc~cm^{-3}}\nonumber\\
\ee
where ${\rm DM_E}= {\rm DM} - {\rm DM_{gal}}=1222.4~{\rm pc~cm^{-3}}$ is the extragalactic DM of FRB 171209 \citep{oslowski19}.

At $z=0.82$, the uncertainty of the IGM DM is $\sigma_{\rm IGM}\sim200~{\rm pc~cm^{-3}}$ \citep{mcquinn14}. Meanwhile, since the host galaxy of GRB 110715A is similar to that of FRB 121102, we take the DM contribution from the interstellar medium (ISM) as ${\rm DM_{ISM}}\lesssim200~{\rm pc~cm^{-3}}$ \citep{sanchez-ramirez17,tendulkar17}. Therefore, even considering the large-scale structure fluctuation and a possible large DM from the ISM, there is still a large DM excess ${\rm DM_{loc}}\sim550~{\rm pc~cm^{-3}}$.
This DM excess is likely contributed by the GRB-associated SN occurred $t\simeq6.4~{\rm yr}$ before FRB 171209.
In the free-expansion phase, the DM provided by a young SNR with mass $M$ and kinetic energy $E_{\rm SN}$ can be estimated as \citep[e.g.,][]{piro16,yangzhang17}
\be
{\rm DM_{SN}}&=&\frac{\eta M^2}{8\pi\mu_mm_pE_{\rm SN}t^2}=630~{\rm pc~cm^{-3}}\nonumber\\
&\times&\eta\left(\frac{M}{M_\odot}\right)^{2}\left(\frac{t}{6.4~{\rm yr}}\right)^{-2}\left(\frac{E_{\rm SN}}{10^{51}~{\rm erg}}\right)^{-1}
\ee
where $\mu_m=1.2$ is the mean molecular weight for a solar composition in the SNR ejecta, and $\eta$ is the ionization fraction of the medium in the SNR. We can see that for a typical SN with a few times solar masses, the corresponding DM contribution could reach the required host-galaxy DM of FRB 171209. One should check the the free-free absorption in the SN.
For a young SNR, the free-free optical depth through the ejecta shell is
\be
\tau_{\rm ff}
&\simeq&(0.018T^{-3/2}Z^2n_en_i\nu^{-2}\bar g_{\rm ff}){\cal L}\simeq3600~\left(\frac{T}{10^4~\unit{K}}\right)^{-3/2}\nonumber\\
&\times&\left(\frac{M}{M_\odot}\right)^{9/2}\left(\frac{E_{\rm SN}}{10^{51}~\unit{erg}}\right)^{-5/2}\left(\frac{t}{1~\unit{yr}}\right)^{-5}\left(\frac{\nu}{1~\unit{GHz}}\right)^{-2},\nonumber\\
\ee
where $n_e$ and $n_i$ are the number densities of electrons and ions, respectively, and $n_e = n_i$ and $Z = 1$ are assumed for an ejecta with a fully ionized hydrogen-dominated composition, ${\cal L}\sim r\sim vt$ is the ejecta thickness, and $\bar g_{\rm ff}\sim1$ is the Gaunt factor. If the SNR ejecta is transparent for FRB, i.e., $\tau_{\rm ff}\lesssim1$, one gets the SNR age \citep[e.g.][]{yang2019}
\be
t\gtrsim5~\unit{yr}\left(\frac{M}{M_\odot}\right)^{9/10}\left(\frac{E_{\rm SN}}{10^{51}~\unit{erg}}\right)^{-1/2},
\ee
where $\nu\sim1~\unit{GHz}$ and $T\sim10^4~\unit{K}$ are taken. 
This is consistent with the $6.4~{\rm yr}$ time delay between FRB 171209 and GRB 110715A.

\section{Summary and Discussions}
\label{sec:con}

Lacking multi-wavelength observational data of FRBs, it is hard to constrain their physical origin. It has been suggested that at least some FRBs may be physically associated with GRBs \citep{zhang14,metzger17}. The GRB may leave behind a long-lived magnetar, which may produce FRBs through ejecting magnetosphere upon collapse \cite{falcke14}, or more likely, produce repeated bursts through crust cracking or magnetic reconnection \citep[e.g.][]{popov10,katz16,beloborodov17,kumar17,yangzhang18,wangw18}.

We searched for possible GRB-FRB associations based on the  localization data of 110 FRBs and the precise localization data of 1440 GRB afterglows.
We found that the long-duration GRB 110715A is within the error box of FRB 171209 and the redshift of the GRB 110715A is lower than the maximum redshift derived from the DM of the FRB 171209.
Taking the factors of spatial location, time of occurrence, and the redshift criterion, we derive a chance probability of 2.3\% to 1.1\%, corresponding to a 2.28$\sigma$ to 2.55$\sigma$ confidence level for the association.

Even though the chance coincidence probability cannot establish a firm association between GRB 110715A and FRB 171209, we nonetheless investigated whether there exists a self-consistent physical picture to make a connection between the two.
We modeled the afterglow of GRB 110715A and identified a shallow-decay signature, which is consistent with energy injection by a millisecond magnetar with $P_0 <  3.59 $ ms and $B_{\rm p}< 4.59 \times 10^{15}$ G.
With the Milky Way and IGM contributions subtracted, the observed DM of FRB 171209 has an excess of $\sim 950$ pc cm $^{-3}$, which is consistent with the DM contribution of a young ($\sim 6.4$ yr old) SNR associated with GRB 110715A with a few solar masses and kinetic energy $E_{\rm SN}\sim 10^{51}$ erg. The requirement that the free-free optical depth
$\tau_{\rm ff}\lesssim1$ suggests that FRBs can be observable only a few years after the explosion, consistent with the observed 6.4 yr delay between GRB 110715A and FRB 171209. FRB 171209 so far does not show a repeating behavior. Its lightcurve shows one single pulse without a noticeable temporal structure \citep{oslowski19}. The intrinsic duration is sub-millisecond. In principle, the burst could be an one-off event. If it is associated with GRB 110715A, it may be related to the collapse of the supramassive neutron star at such a late epoch \citep{falcke14,zhang14}. However, contrived conditions are needed to allow the collapsing time to be at such a late stage after the spindown timescale.
More likely, FRB 171209 may be one of many repeating bursts powered by the magnetar harbored in GRB 110715A \citep{murase16,metzger17}. Searching for repeating bursts from FRB 171209 would be essential to test this possibility.

Observationally, no SN was reported for GRB 110715A \citep{sanchez-ramirez17}. This is not surprising, since GRB 110715A is not nearby and is a high-luminosity long GRB with a bright optical afterglow. The SN signature is likely outshone by the afterglow. It is well known that essentially every long GRB is accompanied by a Type Ic SN \citep{Woosley06}, so that invoking a SN to account for the extra DM from FRB 171209 is justified.

\acknowledgments
XGW, DBL and EWL acknowledges support from the National Natural Science Foundation of China (grant No.11673006, U1938201, 11533003, 11773007), the Guangxi Science Foundation (grant No. 2016GXNSFFA380006, 2017GXNSFBA198206, 2018GXNSFFA281010, 2017AD22006, 2018GXNSFGA281007), the One-Hundred-Talents Program of Guangxi colleges, and High level innovation team and outstanding scholar program in Guangxi colleges. BZ and JWL acknowledge the UNLV Top-Tier Doctoral Graduate Research Assistantship (TTDGRA) grant for support. We also acknowledge the use of public data from the \emph{Swift} data archives and the FRB catalog (http://frbcat.org).



\end{document}